\documentclass[galaxies,article,accept,pdftex,moreauthors]{Definitions/mdpi} 
\firstpage{1} 
\pubvolume{1}
\issuenum{1}
\articlenumber{0}
\pubyear{2022}
\copyrightyear{2022}
\externaleditor{Academic Editor: Martín Guerrero, Noam Soker and Quentin A. Parker} 
\datereceived{30 January 2022} 
\dateaccepted{25 February 2022} 
\datepublished{} 
\hreflink{https://doi.org/} 

\Title{Follow-Up of Extended Shells around B[e] Stars}

\TitleCitation{Follow-Up of Extended Shells around B[e] Stars}

\Author{Tiina Liimets$^{1,}$*\orcidA{}, Michaela Kraus $^{1}$\orcidB{}, Alexei Moiseev $^{2}$, Nicolas Duronea$^{3}$, Lydia Sonia Cidale$^{3,4}$ and Cecilia Fari\~na$^{5,6}$}

\AuthorNames{Tiina Liimets, Michaela Kraus, Alexei Moiseev, Nicolas Duronea, Lydia Sonia Cidale and Cecilia Fari\~na}

\AuthorCitation{Liimets, T.; Kraus, M.; Moiseev, A.; Duronea, N.; Cidale, L.S.; Fari\~na, C.}

\address{%
$^{1}$ \quad Astronomical Institute, Czech Academy of Sciences, Fri\v{c}ova 298, 25165 Ondrejov, Czech Republic; michaela.kraus@asu.cas.cz\\ 
$^{2}$ \quad Special Astrophysical Observatory, Russian Academy of Sciences, 369167 Nizhnii Arkhyz, 
Russia; moisav@gmail.com\\ 
$^{3}$ \quad Instituto de Astrof\'isica de La Plata (CCT La Plata---CONICET, UNLP), Paseo del Bosque S/N, B1900FWA La Plata, Buenos Aires, 
  Argentina; nduronea@fcaglp.unlp.edu.ar (N.D.); lydia@fcaglp.unlp.edu.ar (L.S.C.)\\
$^{4}$ \quad Departamento de Espectroscop\'ia, Facultad de Ciencias Astron\'omicas y Geof\'isicas, Universidad Nacional de La Plata, Paseo del Bosque S/N, B1900FWA La Plata, Buenos Aires, Argentina\\
$^{5}$ \quad Isaac Newton Group, Apartado de Correos 321, 38700 Santa Cruz de La Palma, Canary Islands, Spain; cf@ing.iac.es \\ 
$^{6}$ \quad Instituto de Astrofísica de Canarias, 38205 La Laguna, Tenerife, Spain 
}
\corres{Correspondence: tiina.liimets@asu.cas.cz} 

\abstract{B[e] stars are massive B type emission line stars in different evolutionary stages ranging from pre-main 
sequence to post-main sequence. 
Due to their mass loss and ejection events these objects 
deposit huge amounts of mass and energy into their environment and 
enrich it with chemically processed material, contributing significantly
to the chemical and dynamical evolution of their host galaxies.  
However, the large-scale environments of these enigmatic objects have not attracted much attention.
The first and so far only catalog reporting the detection of extended shells around a sample of 
B[e] stars was an H$\alpha$ imaging survey carried out in the year 2001, and was limited to bright 
targets in the northern hemisphere.
We have recently started a follow-up of those targets to detect possible evolution of their 
nebulae in the plane of the sky over a baseline of two decades. 
Furthermore, we extend our survey to southern targets and fainter northern ones 
to complement and complete our knowledge on 
large-scale ejecta surrounding B[e] stars. 
Besides imaging in H$\alpha$ and selected nebular lines, we utilize 
long-slit and 3D spectral observations across 
the nebulae to derive their physical properties.
We discovered pronounced nebula structures around 15 more objects, 
resulting in a total of 27 B[e] stars with a large-scale nebula. 
Here we present our (preliminary) results for three selected objects: the two massive supergiants MWC137 
and MWC 314, and the unclassified B[e] star MWC 819.
}


\begin{document}



\section{Introduction}
B[e] stars are massive B-type emission-line stars. Their spectra exhibit intense Balmer emission, 
numerous forbidden lines, as well as permitted lines of low ionized metals. 
These stars can be found in different 
evolutionary stages, from pre- to post-main sequence and spreading over a large mass range 
(from sub-solar up to 70\,$M_{\odot}$ for the currently known members\endnote{We would like to emphasize that stars with the B[e] phenomenon are not predicted by any of the currently known stellar evolution theories. This might be related to the fact that these models are purely 1D and do not account for aspherical mass loss, which is typically seen in B[e] stars. Hence this limit of 70\,$M_{\odot}$ might not be real and more massive B[e] stars might exist, which just have not been discovered yet.}). 
In addition, their spectral energy distributions display a strong 
infrared excess due to the hot, 500 to 1000 K, 
circumstellar dust (e.g., \cite{1998A&A...340..117L}). This dust is suggested to be 
located in a ring or disk-like structure based on polarimetric observations. 
The confinement of the dust into a disk has been confirmed by interferometric observations 
of the closest and brightest objects \cite{2007A&A...464...81D,2011A&A...526A.107M,2012A&A...548A..72C}. 
Furthermore, detailed analysis of interferometric and high-resolution spectroscopic observations revealed 
that the stars are surrounded by multiple rings of gas that revolve the central object on 
(quasi-)Keplerian orbits (see, e.g., 
\cite{2016A&A...593A.112K,2018MNRAS.480..320M,2011A&A...526A.107M,2012A&A...548A..72C,2018A&A...612A.113T} 
). The location of the dust can (but must not) coincide with some individual gas rings. 
The distribution of rings around each object is found to be unique and independent on whether the star 
is presumably single or part of a binary system~\cite{2018MNRAS.480..320M}.
The composition and dynamics of these small-scale discs~(several AU), 
in particular around the B[e] supergiants, 
have been studied intensively to unveil their possible formation mechanisms for which critical rotation, 
binary interaction, as well as slow-wind solutions and pulsation triggered ejections are among the 
suggested scenarios (for details see reviews~\cite{2014AdAst2014E..10D,2019Galax...7...83K}).

In addition to the small-scale dusty discs, B[e] stars are surrounded by large-scale 
(up to several pc) ionized gas, 
which has not been studied much. A pioneering 
work \cite{2008A&A...477..193M} presented 
an H$\alpha$ imaging survey for 25 northern B[e] stars, and report the detection of 
extended shells around 50\% of the targets with various shapes: spherical and ring nebulae, 
spiral-arm like features, bipolar and unipolar-lobe structures. This diversity in shapes indicates that 
the formation mechanism is most likely not a unique process. Furthermore, the physical mechanism 
creating such mass ejection is not well known. Therefore, to have a statistically meaningful 
sample of B[e] targets with extended nebulae, and to investigate details of their mass-loss history  
we started an observational campaign with two goals. Firstly, to perform a follow-up of the 
targets with already detected nebulae. Secondly, 
to enlarge the sample of B[e] stars with confirmed nebulae by adding fainter, 
yet unexplored objects located in the north and by extending the survey 
to the southern sky. In this proceedings we present results for selected objects of our project.


\section{Observations and Data Reduction}

Our follow-up H$\alpha$ imaging survey is mostly carried out with the Nordic Optical 
Telescope\endnote{\url{http://www.not.iac.es/} (accessed on 2022, January 31). 
} (NOT) at La Palma and 
with the Danish Telescope in La Silla, for the northern and southern targets, respectively. 
Average exposure times are 15--30 min. For deeper images of selected targets 
GMOS \cite{2004PASP..116..425H,2002PASP..114..892A} 
at Gemini South was used while for very large nebulae we utilized the Isaac Newton Telescope (INT). 
Complementing long-slit spectroscopic observations have been secured already for a 
small sample of objects. These observations extend over the wavelength range from $6350$ to 
$6850$\,\AA \ with a spectral resolution of $R \sim 10\,000$ and have been carried out at 
the NOT as well.
This wavelength range covers nebular emission 
lines---\hbox{H$\alpha$ 6563 \AA,} [NII] $\lambda\lambda$6548,6583, 
[SII] $\lambda\lambda$ 6716,6731. 
In addition, we have exploited the 
Scanning Fabry-Perot Interferometer (FPI) mounted at the Russian 6m Big Telescope Alt-azimuth (BTA). 
More observational details of our survey will be presented in a forthcoming paper (Liimets et al. in prep). 
In this proceedings we will restrict to details about observations 
and data reduction for the targets under discussion. 
These are the two supergiant stars MWC 137, and MWC 314, and the object MWC 819, which currently 
has an unknown evolutionary state.

The H$\alpha$ images of the B[e] supergiant MWC 137 were obtained with the NOT 
and the Danish Telescope in 2016 and 2019, respectively. 
For the spectral observations the long-slit possiblities at the NOT 
as well as the scanning FPI mode at the BTA were exploited. 
Further observational details of the data used to investigate MWC 137 
can be found from \citep{2017AJ....154..186K,2021AJ....162..150K}.

The imaging observations of the large-scale bipolar nebula of MWC 314 have been performed 
with the Wide Field Camera at the INT, which provides a Field of View (FOV) $33'\times33'$ and a 
pixel scale $0''.33$ pix$^{-1}$. 
On 2019 October 3 five dithered frames with an exposure time of 10 min each were acquired 
totalling 50 min. 
The narrowband H$\alpha$ filter No. 196 was used, which includes also the [NII] doublet. 
Standard routines in IRAF\endnote{IRAF is distributed by the National 
Optical Astronomy Observatory, which is operated by the Association 
of Universities for Research in Astronomy (AURA) under cooperative 
agreement with the National Science Foundation.} were used for the data reduction. 
Spectral observations of MWC 314 
were performed at the BTA with the multi-mode focal reducer SCORPIO-2 \cite{2011BaltA..20..363A}, 
which provides FOV $6'\times6'$ with a $0''.4$ pixel scale in the 2 $\times$ 2 binning mode. 
To cover the entire nebula, it had to be split into four fields.
The 3D spectroscopic cubes obtained in the scanning FPI mode have a resolution of R$\sim$16000 in the 
narrow spectral range containing the H$\alpha$ emission line. The observed four fields were 
merged into a mosaic cube with an angular size $18'\times12'$. 
Details of the FPI data reduction can be found in \cite{2021AstBu..76..316M}.

A new deep H$\alpha$ image of MWC 819 was acquired on 2020 November 15 with GMOS attached to Gemini South 
(program ID GS-2020B-Q-210).
The FOV was $5'.5\times5'.5$ with a 
pixel scale $0''.16$ pix$^{-1}$ in the utilized 2x2 binning mode. 
The observations were taken with filter G0336, which covers, 
besides H$\alpha$, also the nebular lines from the [NII] $\lambda\lambda$6548,6583 doublet.
To remove the gaps between the CCDs the total exposure 
time of 10 min was divided into 5 subframes of 2 min each. For the data reduction 
the official Gemini Observatory software DRAGONS \cite{2019ASPC..523..321L} was used.

Finally, we have at our hands the raw H$\alpha$ images from 2001 observed with the 
Mt. Palomar 60 inch telescope. Details about the observations are given in \cite{2008A&A...477..193M}. 
We have reduced the images using standard IRAF routines.

\section{Results and Discussion}


Our H$\alpha$ imaging survey has been completed.
To the already published H$\alpha$ survey
of 25 B[e] stars \citep{2008A&A...477..193M}, we added 32 objects, 
9 in the north and 23 in the southern sky. 
Similarly to the first survey, we report detecting 
large-scale nebular features around 50\% of the observed stars, that is 4 in the north and 11 in the south. 
The full description of our 
survey will be published separately (Liimets et al. in prep), but the
results for five objects from our sample (MWC 137, [GKF2010] MN 83, \hbox{[GKF2010] MN 108,}
[GKF2010] MN 109, and [GKF2010] MN 112) have been presented in 
\cite{2017AJ....154..186K,2020AJ....160..166C,2021AJ....162..150K}.  
In the following, we summarize 
our work on MWC 137 and present preliminary results for two additional objects:  MWC 314 and \hbox{MWC 819.}

\subsection{MWC 137}

The object MWC 137 has been classified as a supergiant star that has just left the main sequence. 
Its mass and age have been estimated to be $37^{+9}_{-5}\,M_{\odot}$ and $4.7\pm 0.8$\,Myr, 
respectively \cite{2021AJ....162..150K}.
Our detailed investigations of the various data sets presented in \citep{2017AJ....154..186K,2021AJ....162..150K} have led to the following picture of the environs of the B[e] supergiant MWC 137.

\begin{figure}[H]
\includegraphics[width=7cm, trim=0 0 0 0]{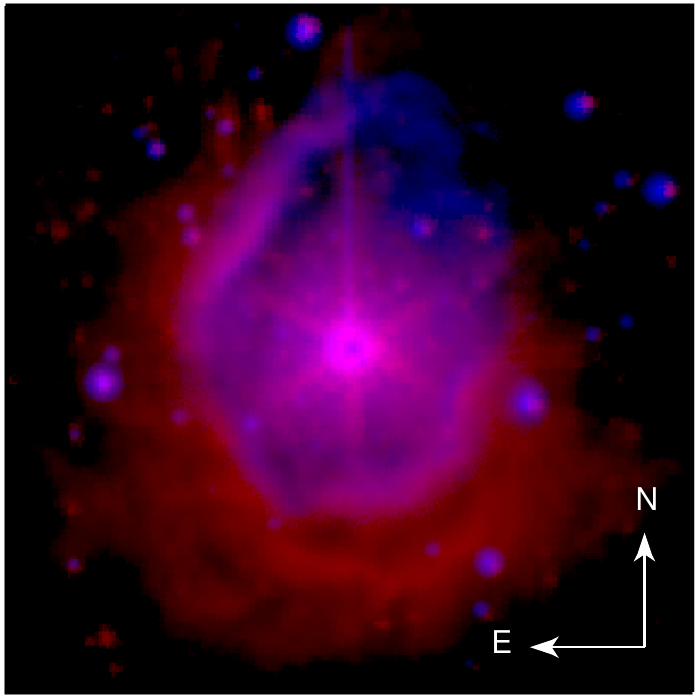}
\caption{Large-scale nebula of MWC 137. ALFOSC H$\alpha$ iamge with blue and 
the \textit{Sptizer} 3.6 $\upmu$m with red. FOV $2'\times2'$. 
Figure taken from \cite{2017AJ....154..186K} (their Figure 10, 
$\copyright$ AAS, reproduced with permission).}
\label{F-137} 

\end{figure}

The central star is surrounded by hot CO gas forming a subarcsecond size 
disc displaying Keplerian rotation.
Further out from the center a prominent 
arcminute scale ionized nebula is visible (Figure~\ref{F-137}). In the Figure, 
the emission from the H$\alpha$+[NII] doublet is depicted in blue, while with red we 
present the warm and cool dust observed with the Spitzer Space Telescope in the 3.6 $\upmu$m band. 
The dust visible in the infrared has a slightly larger size than the 
optical nebula, appart from the north-west (NW) direction, where the warm and cool dust 
seems to be absent (see Figure~\ref{F-137}). The ionized gas has an elliptical shape with varying intensity 
areas, presenting several brigther filaments. We compared our newly obtained 
H$\alpha$ image from 2019 with the image from 2001 \citep{2008A&A...477..193M}. 
Our baseline of 18.1 years did not reveal any morphological changes. Neither 
could we yet detect any expansion in the plane of the sky.

\begin{figure}[H]
\includegraphics[width=5.3cm, trim=0 0 0 0]{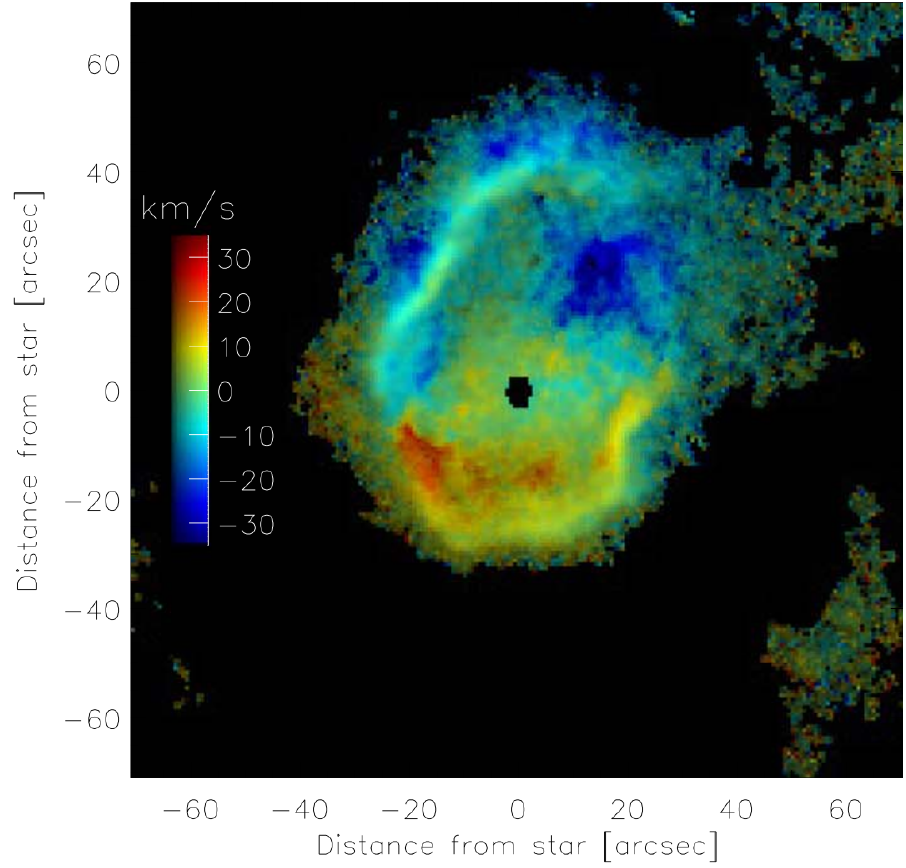}
\includegraphics[width=6.6cm, trim=0 0 0 0]{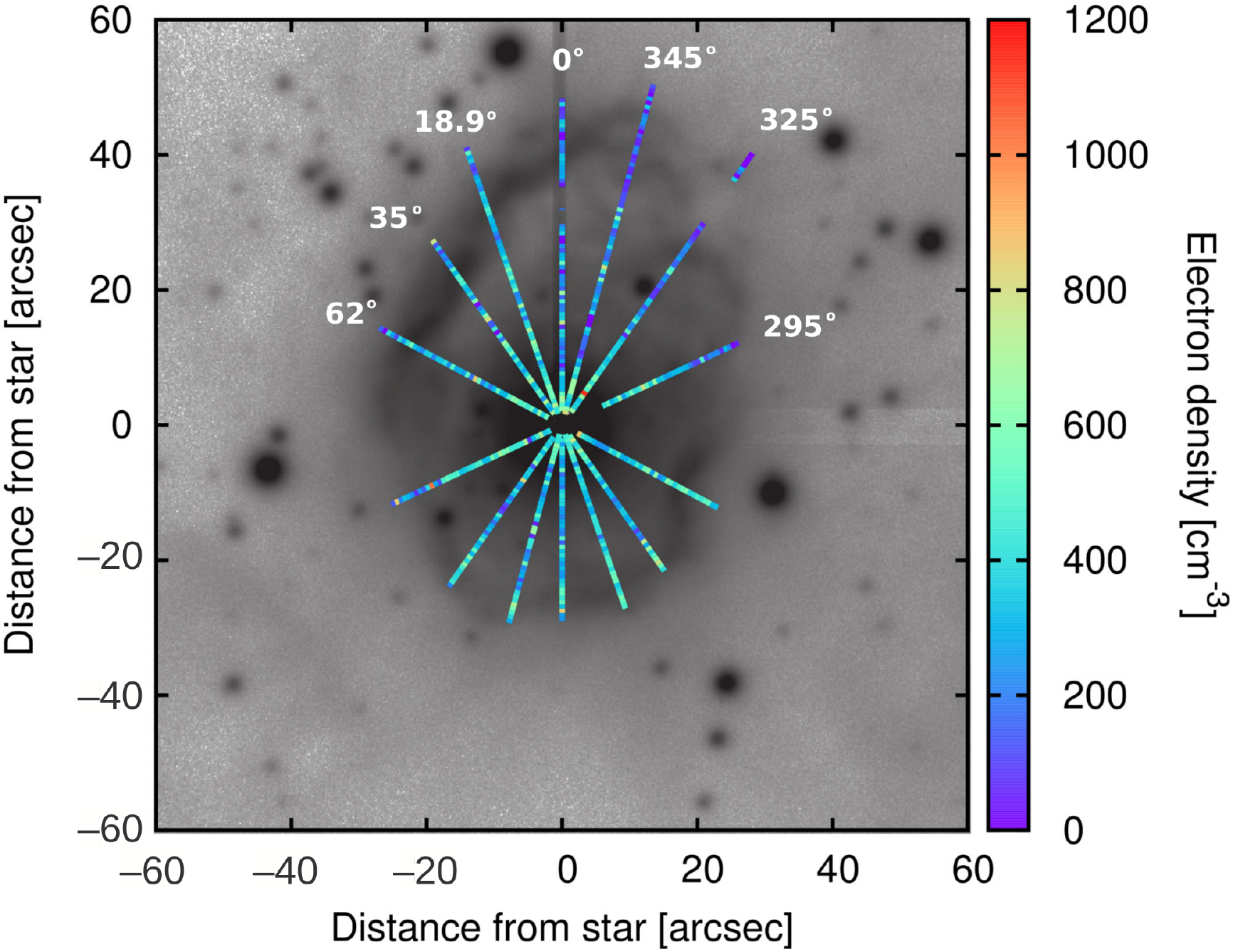}
\caption{Radial velocity measurements of the MWC 137 from the FPI observations (\textbf{left}) 
and electron density distribution across the nebula obtained from the ratio of the 
[SII] $\lambda\lambda$ 6716,6731 lines measured in 
the NOT+ALFOSC long-slit spectra (\textbf{right}). On both panels north is up and east to the left. 
Figures taken from \citep{2021AJ....162..150K} (their Figure 4a, left panel, and Figure 9, right panel, 
$\copyright$ AAS, reproduced with~permission).}
\label{F-137velden} 
\end{figure}

We further investigated the ionzied gas in the optical range using our various spectral observations. 
In Figure~\ref{F-137velden} left we present our radial velocity (RV) measurements of the 
combined [SII] $\lambda\lambda$ 6716,6731 doublet obtained with the scanning FPI  
which is providing spectral coverage over the whole FOV. The RV pattern is 
complex, which is also supported with our long-slit spectra 
obtained with the NOT+ALFOSC\endnote{Alhambra Faint Object Spectrograph and Camera \url{http://www.not.iac.es/instruments/alfosc/} (accessed on 2022, January 31).} 
at various position angles across the nebula 
(see Figure 4 in \cite{2017AJ....154..186K} and Figure 5 in \cite{2021AJ....162..150K}). 
However, the northern part is mostly 
blue shifted and the southern part red shifted. In addition, from our long-slit spectra 
we could estimate electron densities utilizing the same [SII] doublet and assuming a 
nebular temperature of 10,000 K. 
Similarly to the RV, the electron densities show a varying pattern across
the nebula, from 0 to 800 cm$^{-3}$ (Figure~\ref{F-137velden} right). 
The higher densities are at the regions of intense 
emission and at the central area. 

\vspace{-6pt}

Moving to even larger distances from the center of MWC 137 
we find that the optical nebula is surrounded by a vast amount of cold molecular gas 
(see Figure 12 in \cite{2017AJ....154..186K}) detected with the Atacama Pathfinder EXperiment (APEX). 
The various molecular clouds resolved in the radio 
regime have masses from 15 to 245 M$_{\odot}$  and seem to embrace the ionized nebula in the west, 
south, and east, clearly implying an interstellar origin rather 
than resulting from the evolution of the central star.

\subsection{MWC 314}

\begin{figure}[H]
\includegraphics[width=13.5cm]{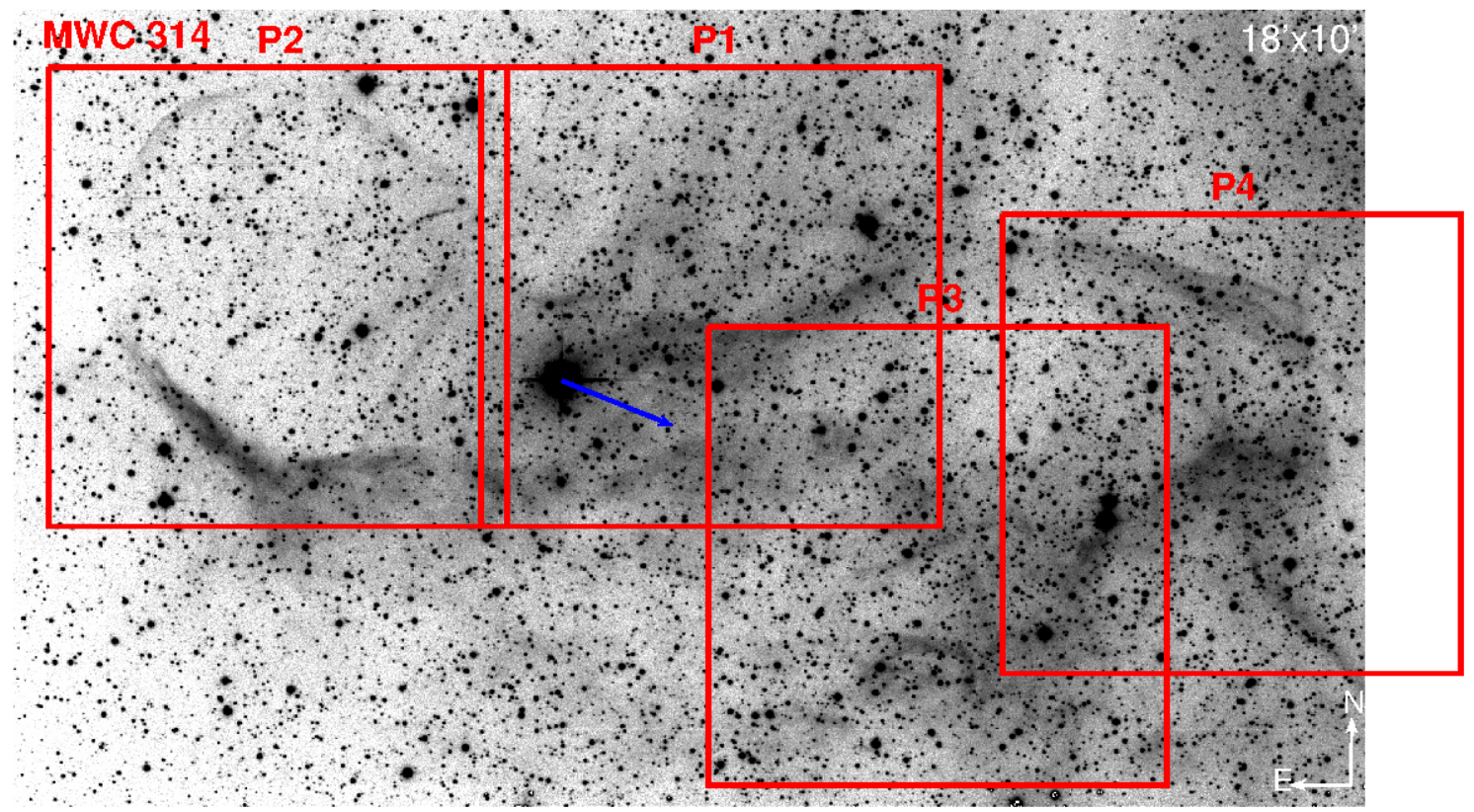}
\caption{H$\alpha$ image (INT) of the bipolar nebula around MWC 314. 
The pointings for the spectroscopic FPI observations are shown in red. 
See text for more details.}
\label{F-314}
\end{figure}

MWC 314 is one of the most luminous stars in the Galaxy \citep{1998A&AS..131..469M}. 
The debate over its nature is ongoing---a Luminous Blue Variable (LBV, e.g.,~\citep{2006PASP..118..820W}) 
or a B[e] supergiant (e.g.,~\citep{2016A&A...585A..60F}). 
It seems to be established though that the central object is a contact binary. 
But the masses of the system and of the individual stars are also controversial. 
Suggested are a total mass of the system of $66\pm 9\,M_{\odot}$ \cite{2013A&A...559A..16L} and of
$\sim$20 $M_{\odot}$ \cite{2016MNRAS.455..244R}. This discrepancy renders it difficult to assign the 
object a proper evolutionary state.
MWC 314 is surrounded by a huge bipolar nebula extending 18' across (\citep{2008A&A...477..193M}, 
Figure~\ref{F-314}), which considering the 
distance of 3 kpc \cite{1998A&AS..131..469M} translates into a physical size larger than 13 pc.

We started the analysis with our H$\alpha$ image from 2019, which we compared to the one
from 2001 to find any morphological changes and/or expansion in the plane of the sky. 
Comparing the bipolar lobes in the east-west direction, no  morphological changes were detected. 
For the expansion in the plane of the sky, we exploited the so-called 
magnification method (see e.g., \citep{1999AJ....118.2430R,2018A&A...612A.118L}), 
which is able to detect expansion down to 0.1 pixels on the residual image that has 
been created from the images taken at two different dates. To properly apply the 
magnification method, further processing of the images was needed,  
which included pixel by pixel, seeing, and flux matching. 
Further details of this processing can be found from \cite{2021AJ....162..150K,2018A&A...612A.118L}. Unfortunately, our baseline of 18 years did not reveal 
any expansion in the plane of the sky. 

Next, from inspection of our deep image, a possible third lobe NW of the central object 
might be identified, which is barely seen on the image from 2001 \cite{2008A&A...477..193M}.
To further investigate this lobe, we included it into our spectroscopic observations with the FPI.
It is almost fully covered by the pointing P1  
(see Figure~\ref{F-314}). 
The other pointings were chosen such that they cover most of the north-east (NE) 
lobe (P2) and the lobe in the west (W, P3 and P4).
The proper motion of the central star, according to Gaia DR2 \citep{2018yCat.1345....0G}, 
is \hbox{$-2.3$ mas yr$^{-1}$} and $-4.8$ mas yr$^{-1}$ in Right Ascension and Declination, respectively. 
A blue arrow, indicating the direction of this movement, is drawn in Figure~\ref{F-314}, 
which fits well with a plausible swept-back nature of the bipolar nebula 
similar to the swept-back nebula detected around the classical nova GK Persei 1901 
(see Figure 2 in \citep{2004ApJ...600L..63B}). This could also mean that the western lobe, 
which has been previously considered in the literature as the plausible counterpart to the NE lobe, 
must not necessarily be connected to \hbox{MWC 314}. In fact, the NE lobe appears to be rather 
extended towards the west and south, but these features have  no clear counterparts in NE direction. 

\begin{figure}[H]
\includegraphics[width=12.0cm]{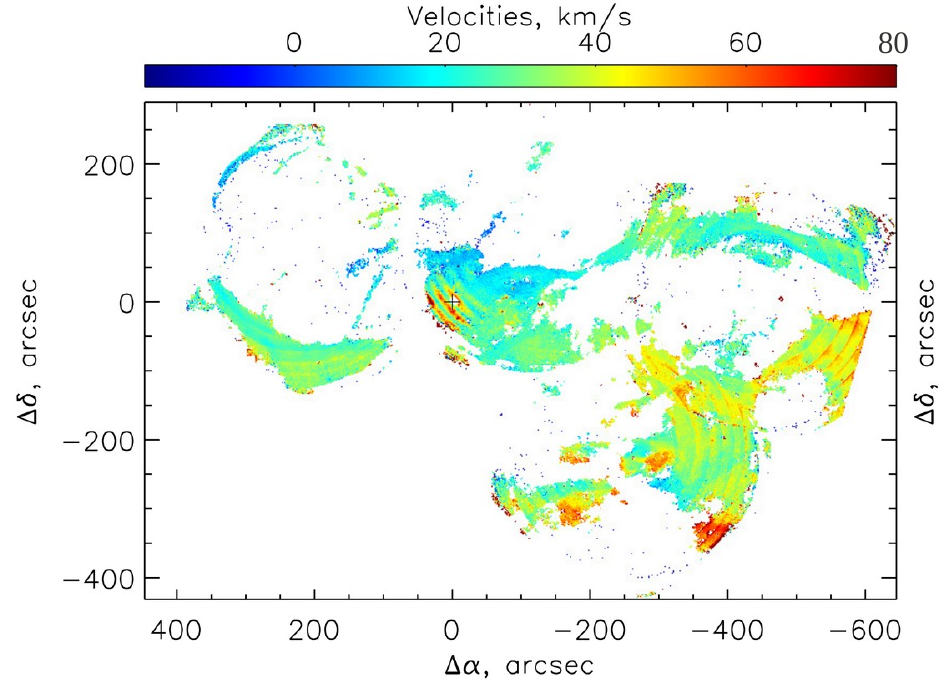}
\caption{Velocity field of the bipolar nebula around MWC 314 measured from the [SII] lines observed with the FPI.}

\label{F-314vel}
\end{figure}

To test whether the NW structure is the  actual counterpart to the NE lobe, 
velocity measurements over the entire nebula are required. 
Unfortunately, the 3D spectroscopic observation (see Figure~\ref{F-314vel}) had no signal in that area. 
However, as pointed out by Bruce Balick during the conference, our proposed new 
lobe in the NW seems to be replenished with gas, while the two previously considered lobes are not. 
This is, so far, the only argument speaking against the scenario of the new lobe to be related 
to the MWC 314. Without further observations, in particular a repetition of the velocity measurements, 
we cannot confirm yet the possible connection of the NW lobe with MWC 314.

Thirdly, we analysed the velocity field of the lobes of MWC 314 which is 
presented in Figure~\ref{F-314vel}.
As was already mentioned, the signal-to-noise was too low to get any measurements 
at the position of the NW lobe. In addition, due to problems with 
subtraction of the bright airglow emission lines, the RV 
measurements suffer from artefacts, 
which appear in Figure~\ref{F-314vel} as concentric rings with artificially higher radial velocities.
Further analysis using new 
data reduction algorithms is required to 
remove these spurious signals. Nevertheless, from the 
RV figure we can tentatively conclude that the northern parts are predominantly 
blue shifted while the southern parts are redder, broadly from $-15$ km s$^{-1}$ 
to $+40$ km s$^{-1}$, respectively.


\subsection{MWC 819}

\begin{figure}[H]
\includegraphics[width=13.5cm]{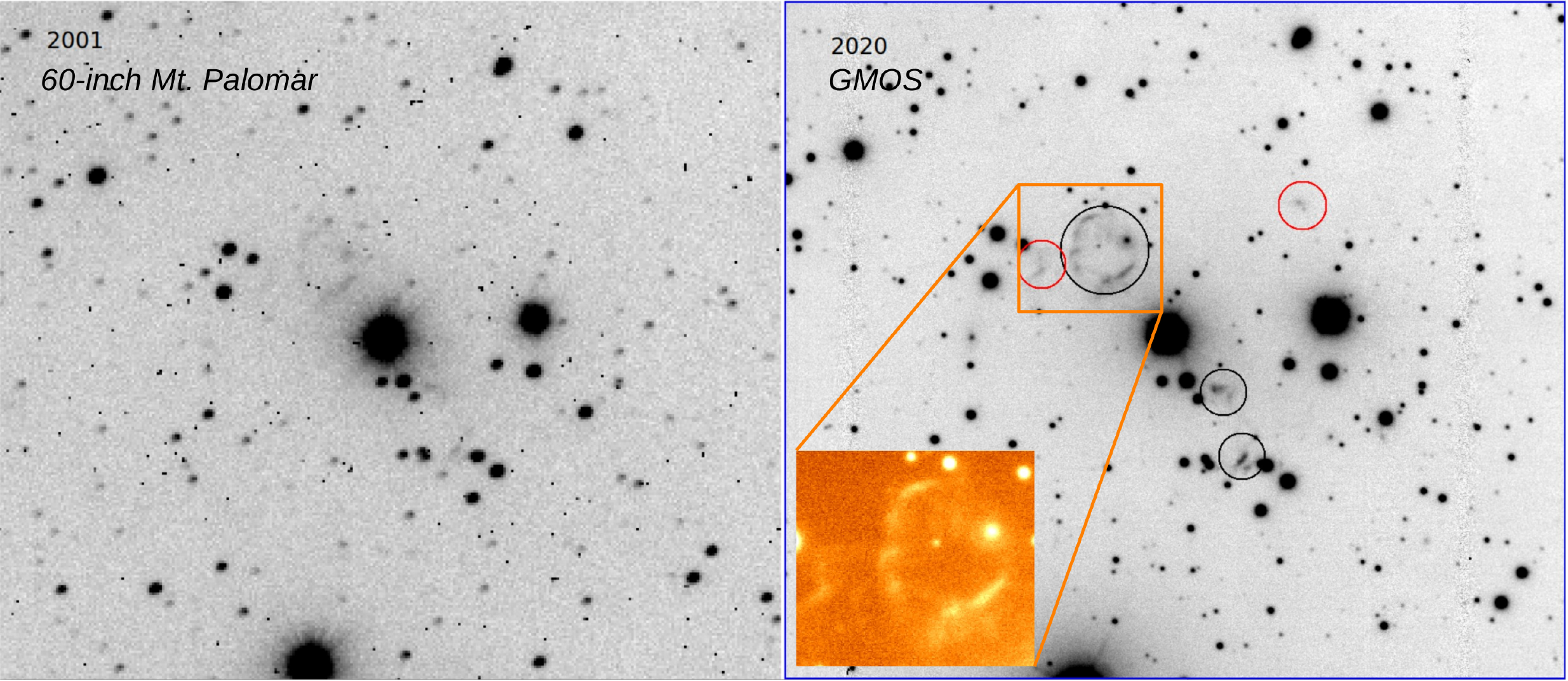}
\caption{Nebular features of the B[e] star MWC 819 seen in 2001 from \cite{2008A&A...477..193M} (\textbf{left}) 
and in our new, deeper image taken in 2020 (\textbf{right}). North is up, east is left.}
\label{F-819}
\end{figure}

MWC 819 has long been listed as an unclassified B[e] star, but recently, \cite{2018PASP..130k4201A} 
concluded from their analysis of new optical and infrared data that it might be a compact 
protoplanetary nebula. 
Consequently, the central object should currently be evolving into a white dwarf.
On the left side panel of Figure~\ref{F-819} we present 
the image from 2001 \cite{2008A&A...477..193M} and on the right side panel we show our H$\alpha$ 
image from 2020. The FOV of each image is $3'.6\times3'.1$ and \hbox{MWC 819} is the bright source in 
the center of each image.

From the 2001 data MWC 819 was identified as having a unipolar nebula \cite{2008A&A...477..193M}, 
visible as the faint roundish nebular feature in the NE. 
Our new and considerably deeper image resolves the previously detected features in more detail and 
reveals an additional possibly related nebular ejecta. In particular, 
the "unipolar lobe'' (marked with a big black circle in our 2020 image) 
for which a zoom-in is presented with an orange square 
(FOV $38''\times36''$), resembles an individual nebula around a faint, 
possibly background object that was not detected in the 2001 image.
To confirm this hypothesis, additional observations and analyses are needed. 
We would also like to draw the readers attention to the smaller nebular features 
in the south-west (SW) marked with two smaller black circles. Those features were not mentioned 
in \cite{2008A&A...477..193M}. However, after being clearly resolved in our deeper image, 
faint counterparts can be detected also in the 2001 image. Furthermore, with red we mark 
previously not detected features. One in the NE and the other in the NW direction. 
The latter is especially interesting, because due to its position perpendicular  
to the main nebular features in the NE to SW direction, it 
might be possible that a jet is emanating from MWC 819. In fact, jets have been discovered in a few 
other B[e] stars (e.g., MWC 137 \cite{2016A&A...585A..81M,2017AJ....154..186K}, 
MWC 922 \cite{2019MNRAS.484.4529B}, Z CMa \cite{2016A&A...593L..13A}). 

\section{Conclusions}

We have performed a new H$\alpha$ imaging survey of 32 B[e] stars. Of these, 
9 targets are located in the northern hemisphere, and 23 in the southern one. 
Moreover, we re-observed the 25 B[e] stars, whose environment had been imaged 
about two decades ago \cite{2008A&A...477..193M}. 
Our study has two main goals: (i) to search for indications of large-scale 
ejecta around the yet unstudied objects, and (ii) to investigate the morphology 
of the nebulae and their expansion dynamics. Furthermore, all objects with 
detected circumstellar nebulae are observed spectroscopically in a systematic 
way to derive the physical parameters such as temperature and density distribution 
across the nebulae together with their ionization structure and chemical composition. 
The combined information will be used to address the issue of the yet unknown 
mass-loss history of these enigmatic objects. 
Moreover, we expect that the detailed information about the nebulae gained 
during our project will help to settle the yet unknown evolutionary state 
of many of the objects. 

Our survey of the newly imaged objects has revealed that about 50\% (15 objects out of 32) 
of the B[e] stars have detectable nebular features. This detection rate is similar to 
the previously reported one (12 out of 25 objects, \cite{2008A&A...477..193M}). 
The shapes of nebulae display a high diversity, and a full catalog with all 
the images is in preparation. 

While for most objects the analysis is still ongoing and, in particular, 
a wealth of spectroscopic data still needs to be acquired, 
we have presented our (preliminary) results for three objects. 
Two of them (MWC 137 and MWC 314) had been previously imaged, 
but our current baseline of about 18 years provided neither evidence 
for changes in nebular morphology nor indications for nebular expansions. 
However, both objects have rather large distances (about 5 and 3 kpc, respectively), 
so that our current spatial resolution is still too coarse to expect 
detectable changes, given the moderate radial velocities measured in 
both nebulae. Our deep image of the presumably bipolar nebula around 
MWC 314 pointed to a third, faint lobe filled with emission, 
which would fit better to the direction of the proper motion of the star. 
Whether it can indeed be associated with MWC 314 requires better 
radial velocity measurements of the entire nebular structure than the one we could secure so far.

For the third object presented, MWC 819, we have also an old set of recorded data at hand. 
Our new, significantly deeper image suggests that one of the previously detected 
nebular features has been mistakenly identified as uni-polar lobe connected to \hbox{MWC 819}. 
It seems that this feature could be associated instead
to a faint back-ground source, which was not visible on the old image.
However, our new, high-quality image displays other, yet unreported signs for 
ejected matter, maybe in relation with a jet. Confirmation is still pending and 
requires high-quality spectra, which are currently under way, to determine 
the velocity structure of the detected features.

\vspace{6pt} 



\authorcontributions{
Conceptualization and Project administration, T.L., M.K. and L.C.; 
Resources and Data curation: T.L., A.M., N.U.D., L.C. and C.F.;
Formal analysis, Investigation, Software, Methodology and Validation, T.L., M.K., A.M. and N.U.D.; 
Supervision, M.K.;
Visualization, T.L., M.K. and A.M.; 
writing---original draft preparation, T.L. and M.K.; 
writing---review and editing, all; 
Funding acquisition, M.K. and L.C.
All authors have read and agreed to the published version of the~manuscript.}

\funding{This research was funded by the Czech Science foundation (GA \v{C}R, grant number 20-00150S), by CONICET (PIP 0177, PIP 2893) and by the Agencia Nacional de Promoci\'{o}n Cient\'{i}fica y Tecnol\'{o}gica (PICT 2016-1971). The Astronomical Institute of the Czech Academy of Sciences is supported by the project RVO:67985815. This project has also received funding from the European Union's Framework Programme for Research and Innovation Horizon 2020 (2014-2020) under the Marie Sk\l{}odowska-Curie Grant Agreement No. 823734. The analysis of SCORPIO-2 data was carried out within the framework of the government contract of SAO RAS approved by the Ministry of Science and Higher Education of the Russian
Federation.}

\institutionalreview{Not applicable}

\informedconsent{Not applicable}

\dataavailability{Most of the data presented in these proceedings can be acquired using 
the following public archives: 
NOT FITS Header Archive\endnote{\url{http://www.not.iac.es/observing/forms/fitsarchive/} (accessed on 2022, January 31).} for NOT data, 
Isaac Newton Group Archive\endnote{\url{http://casu.ast.cam.ac.uk/casuadc/ingarch/query} (accessed on 2022, January 31).} for INT data,
Gemini Observatory Archive\endnote{\url{https://archive.gemini.edu/searchform} (accessed on 2022, January 31).} for GMOS data,
ESO Archive\endnote{\url{http://archive.eso.org/eso/eso\_archive\_main.html} (accessed on 2002, January 31).} for APEX data, 
and The General Observation Archive\endnote{\url{https://www.sao.ru/oasis/cgi-bin/fetch?lang=en} (accessed on 2022, January 31).} 
for BTA data. Observations obtained with the Danish 
and Mt. Palomar 60 inch telescope can be obtained from TL on a reasonable request. 
}

\acknowledgments{
We thank the anonymous referees for their constructive comments that helped to improve the manuscript.
This research made use of the NASA Astrophysics Data System (ADS) and of
the SIMBAD database, operated at CDS, Strasbourg, France.  
This publication is based on observations obtained at the international Gemini Observatory, a program of NSF’s NOIRLab 
[processed using DRAGONS (Data Reduction for Astronomy from Gemini Observatory North and South)], which is managed by the Association of Universities for Research in Astronomy (AURA) under a cooperative agreement with the National Science Foundation on behalf of the Gemini Observatory partnership: the National Science Foundation (United States), National Research Council (Canada), Agencia Nacional de Investigaci\'{o}n y Desarrollo (Chile), Ministerio de Ciencia, Tecnolog\'{i}a e Innovaci\'{o}n (Argentina), Minist\'{e}rio da Ci\^{e}ncia, Tecnologia, Inova\c{c}\~{o}es e Comunica\c{c}\~{o}es (Brazil), and Korea Astronomy and Space Science Institute (Republic of Korea) under program ID GS-2020B-Q-210 . 
Partially based on observations made with the Nordic Optical Telescope, owned in collaboration by the University of Turku and Aarhus University, and operated jointly by Aarhus University, the University of Turku and the University of Oslo, representing Denmark, Finland and Norway, the University of Iceland and Stockholm University at the Observatorio del Roque de los Muchachos, La Palma, Spain, of the Instituto de Astrofisica de Canarias. 
The data presented here were obtained in part with ALFOSC, which is provided by the Instituto de Astrofisica de Andalucia (IAA) under a joint agreement with the University of Copenhagen and NOT.
Observations on the 6-m telescope of SAO RAS are supported by the Ministry of Science and
Higher Education of the Russian Federation (including contract No. 05.619.21.0016, unique project identifier RFMEFI61919X0016). The Isaac Newton Telescope is operated on the island of La Palma by the Isaac Newton Group of Telescopes in the Spanish Observatorio del Roque de los Muchachos of the Instituto de Astrofísica de Canarias.
We thank Bruce Balick for a valuable discussions on the nebula of MWC 314. }

\conflictsofinterest{The authors declare no conflict of interest.} 



\abbreviations{Abbreviations}{
The following abbreviations are used in this manuscript:\\

\noindent 
\begin{tabular}{@{}ll}
NOT & Nordic Optical Telescope\\
INT & Isaac Newton Telescope\\
BTA  & Big Telescope Alt-azimuth\\
APEX & Atacama Pathfinder EXperiment \\
FPI & Scanning Fabry-Perot Interferometer\\
ALFOSC & Alhambra Faint Object Spectrograph and Camera \\
FOV & Field of View \\
RV & radial velocity \\
N & north\\
S & south\\
E & east \\
W & west\\
NE & north-east\\
NW & north-west\\
SW & south-west\\
LBV & Luminous Blue Variable 
\end{tabular}}

\begin{adjustwidth}{-\extralength}{0cm}
\printendnotes[custom]
\reftitle{References}

\end{adjustwidth}

\end{document}